\documentclass[conference]{IEEEtran}
\IEEEoverridecommandlockouts
\usepackage{cite}
\usepackage{amsmath,amssymb,amsfonts}
\usepackage{algorithmic}
\usepackage{graphicx}
\usepackage{textcomp}
\usepackage{xcolor}
\usepackage{array}
\usepackage{multirow}
\usepackage{float}
\usepackage{hyperref}

\usepackage{xcolor}
\def\BibTeX{{\rm B\kern-.05em{\sc i\kern-.025em b}\kern-.08em
    T\kern-.1667em\lower.7ex\hbox{E}\kern-.125emX}}

\newcommand{\specialcell}[2][c]{%
  \begin{tabular}[#1]{@{}c@{}}#2\end{tabular}}

\makeatletter



\title{Retinal Image Restoration using Transformer and Cycle-Consistent Generative Adversarial Network}

\begin{document}

\author{\IEEEauthorblockN{Alnur Alimanov}
\IEEEauthorblockA{\textit{Department of Computer Engineering} \\
\textit{Bahcesehir University, Istanbul, Turkey}\\
Email: alnur.alimanov@bahcesehir.edu.tr}
\and
\IEEEauthorblockN{Md Baharul Islam}
\IEEEauthorblockA{\textit{Bahcesehir University, Istanbul, Turkey} \\
\textit{American University of Malta}\\
ORCID: 0000-0002-9928-5776}
}


\maketitle

\begin{abstract}
Medical imaging plays a significant role in detecting and treating various diseases. However, these images often happen to be of too poor quality, leading to decreased efficiency, extra expenses, and even incorrect diagnoses. Therefore, we propose a retinal image enhancement method using a vision transformer and convolutional neural network. It builds a cycle-consistent generative adversarial network that relies on unpaired datasets. It consists of two generators that translate images from one domain to another (e.g., low- to high-quality and vice versa), playing an adversarial game with two discriminators. Generators produce indistinguishable images for discriminators that predict the original images from generated ones. Generators are a combination of vision transformer (ViT) encoder and convolutional neural network (CNN) decoder. Discriminators include traditional CNN encoders. The resulting improved images have been tested quantitatively using such evaluation metrics as peak signal-to-noise ratio (PSNR), structural similarity index measure (SSIM), and qualitatively, i.e., vessel segmentation. The proposed method successfully reduces the adverse effects of blurring, noise, illumination disturbances, and color distortions while significantly preserving structural and color information. Experimental results show the superiority of the proposed method. Our testing PSNR is 31.138 dB for the first and 27.798 dB for the second dataset. Testing SSIM is 0.919 and 0.904, respectively. The code is available at \textcolor{magenta}{\url{https://github.com/AAleka/Transformer-Cycle-GAN}}
\end{abstract}


\section{Introduction}
The retina of the central nervous system is responsible for transforming the incoming light into a neural signal and sending it to the brain's visual cortex for processing. Therefore, ophthalmologists can diagnose eye, brain, and blood circulation diseases by analyzing the retinal images. However, sometimes due to poor imaging device configuration, misoperations during the imaging process, and patient restlessness (primarily applicable to children), these images suffer from such degradation types as out-of-focus blurring, image noise, low, high, and uneven illuminations, and color distortion. It suffers a tremendous negative impact on treatment efficiency, the extra time and money expenditure, and even misdiagnosis.

In recent years, deep learning-based techniques \cite{grassmann2018deep, gao2018diagnosis, peng2019deepseenet} are used to detect and classify diseases using retinal images. However, these methods rely on high-quality images during analysis, which is not always possible to acquire in some cases. Researchers also propose traditional methods \cite{shamsudeen2016enhancement, zhang2022double} to enhance these images. However, they often suffer from generalization issues, meaning they cannot be applied to all cases. For example, Zhang et al. \cite{zhang2022double}, took the reflective feature of the fundus camera into account. They proposed a double-pass fundus reflection model (DPFR) that improves the contrast of retinal images. However, the presence of image artifacts is noticeable. Many learning-based techniques \cite{you2019fundus, li2020quality, perez2020conditional, alimanov2022retinal, wan2021retinal, li2021restoration} have been introduced to solve this problem. However, most of these works solved the missing paired data problem by manually degrading original high-quality fundus images. Manually degraded images are different from original low-quality retinal images. Thus, these methods show poor performance when enhancing real degradation effects. For instance, the authors of \cite{li2021restoration} developed a restoration deep learning network that was trained using simulated cataract-like images. However, it performs worse when it comes to restoring real low-quality images. 

\begin{figure*}[!ht]
    \centering
    \includegraphics[width=0.78\textwidth]{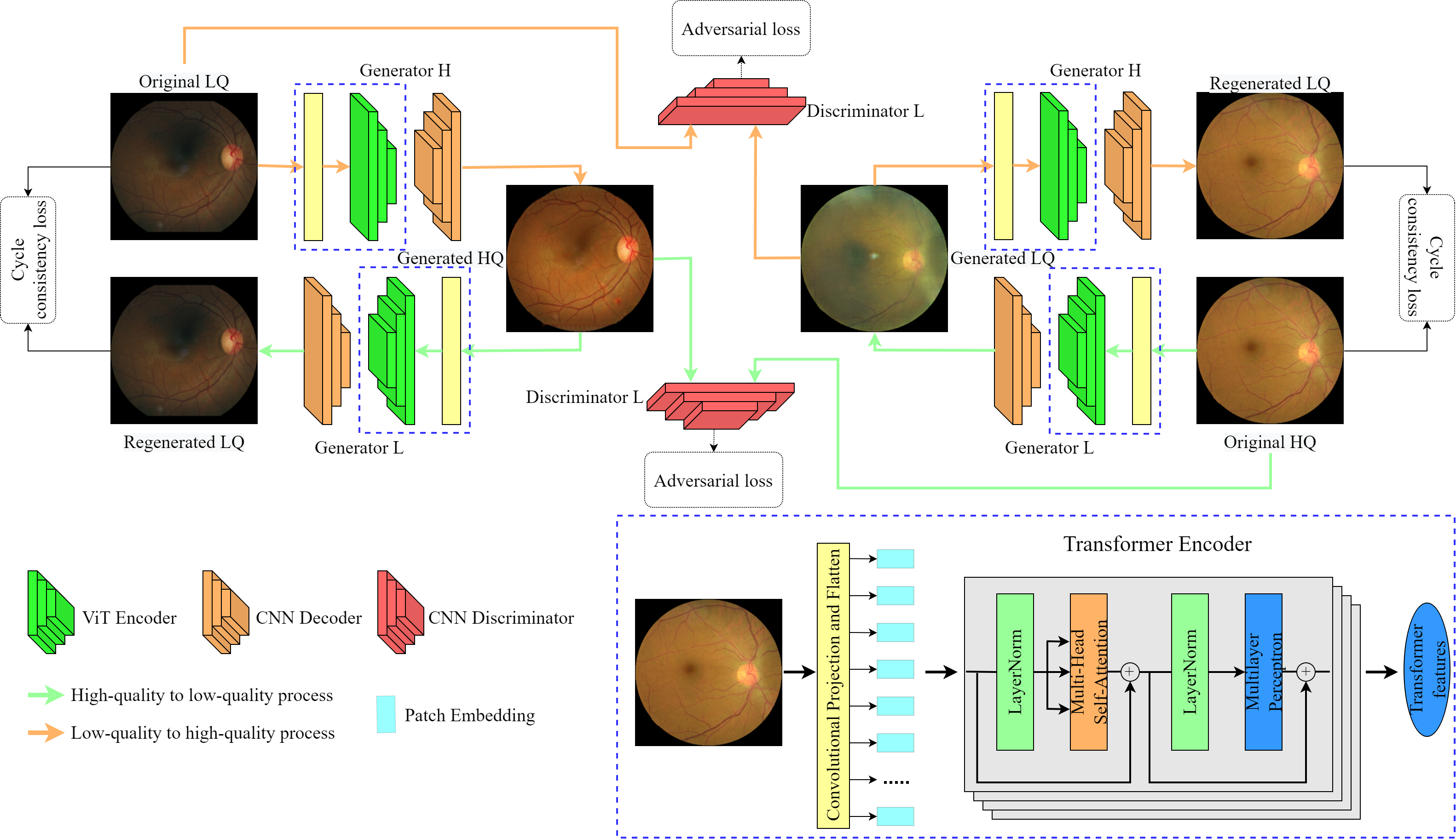}
    \caption{The workflow of the proposed method. Low-quality images go through Generator H and the result - Generated HQ - is inputted into Generator H and Discriminator H. Next, the Generated LQ image is made from high-quality image using Generator L, then this generated image is passed to both Generator H and Discriminator L. After these steps, losses are calculated.}
    \label{fig:architecture}
\end{figure*}

This paper introduces a novel deep learning retinal image restoration model based on a cycle-consistent generative adversarial network with modified generators that do not require paired datasets. Our model replaced the traditional CNN encoder with a vision transformer encoder, resulting in faster convergence, superior quantitative and qualitative outcomes, and better structural and color preservation than other methods. The method has been trained and validated using two publicly available datasets \cite{fu2019evaluation, yoo2020cyclegan}. The contributions can be summarized as follows:

\begin{itemize}
    \item We propose a novel CycleGAN architecture consisting of a vision transformer and convolutional neural network. To the best of our knowledge, this is the first attempt to combine CNN with a transformer in CycleGAN architecture.
    \item Replacing the transformer decoder with a CNN decoder reduced the computational expense compared to pure transformer CycleGAN architecture, which allowed us to use a large image size.
\end{itemize}

\section{Methodology}

\subsection{Model Architecture}
Our method is based on CycleGAN, initially proposed by \cite{zhu2017unpaired}. However, we replaced the CNN encoder in the generator network with a vision transformer encoder, as shown in Fig. \ref{fig:architecture}. Using CNN as a decoder decreased computational expenses compared to a pure transformer network which allowed to increase the image size by a factor of 4, from 64$\times$64 to 256$\times$256. This network consists of a pair of generator networks ($G_H$ and $G_L$) and a pair of discriminators ($D_H$ and $D_L$). The task of each generator is to translate images from one domain to another and to fool the corresponding discriminator. For instance, $G_H$ is trying to generate realistic, high-quality images from original low-quality ones, and $D_H$ receives this generated image and one original high-quality image and calculates the probability of a given image being authentic. As a result, generators and discriminators are playing an adversarial game.

A discriminator is a CNN-based classifier that consists of such layers as 2D convolution, LeakyReLU activation function, and instance normalization. Firstly, a colorful 256$\times$256 image is fed into the convolutional layer with output channels, kernels size, stride, and padding equal to 64, 4, 2, and 1, respectively. Then, we used the LeakyReLU activation function. The result sends to three consecutive downsampling blocks: convolutional layer, instance normalization, and LeakyReLU. Kernel size and padding are equal to 4 and 1, but output features double after each downsampling process starting from 64 and ending with 512. Stride is 1 in the first and 2 in the last two blocks. In the end, it is fed into the final convolutional layer with output channel, stride, and padding equal to 1 and kernel size being 4. As a result, we get a 30$\times$30 patch and make predictions using the sigmoid activation function.

The generator consists of a vision transformer encoder and a CNN decoder. Firstly, each input image of size 256$\times$256 is divided into patches of size 8$\times$8 using a convolutional projection proposed by \cite{wu2021cvt}. It provides additional efficiency. Instead of a depth-wise convolutional layer with batch normalization, we only used a convolutional layer with 1024 output channels, kernel, and stride sizes of 1. Then we flattened the 3D output to 2D and transposed it. It reduced the computation time without affecting the accuracy of the method. Then these flattened sequences are fed into a standard transformer encoder network as shown in Fig. \ref{fig:architecture}. Our network's depth and projection dimensions are 7 and 1024, respectively. The output shape of this encoder is 1024$\times$1024. Therefore, we reshaped it into 1024$\times$32$\times$32 to feed into the CNN decoder. It consists of 3 upsampling blocks: transpose convolution, instance normalization, and ReLU activation function. The output channels have been decreased by 2 after each block; kernel size, stride, padding, and output padding in all blocks are 3, 2, 1, and 1, respectively. The resulted matrix of shape 128$\times$256$\times$256 is converted into a color image using the last convolutional layer with a kernel size of 7, the stride of 1, and padding of 3. The final output is activated using a hyperbolic tangent.

\subsection{Loss functions}


\subsubsection{Adversarial Loss}
Adversarial loss has been developed in the first GAN work \cite{goodfellow2014generative} that  can be formulated as discriminator loss. Given low-quality image $L$, high-quality image $H$, a generator $G_H$ that transforms $L$ to high quality $\hat{H}$, $G_L$ that degrades images (from $Y$ to $\hat{X}$), $D_H$ and $D_L$ classify high- ($H$, $\hat{H}$) and low-quality images ($L$ and $\hat{L}$), respectively. Mathematically, this loss function is expressed as follows:

\begin{align}
    L_{A}(D_L, L, \hat{L}) &= E_{L}[log(D_L(L))] + E_{\hat{L}}[log(1 - D_L(\hat{L}))] \nonumber \\
    L_{A}(D_H, H, \hat{H}) &= E_{H}[log(D_H(H))] + E_{\hat{H}}[log(1 - D_H(\hat{H}))] 
\end{align}

\subsubsection{Cycle Consistency Loss}
Cycle-consistency can be formulated as follows: a generated image $\hat{H}$ translated back to $\acute{L}$ should not be different from $L$. The same things apply for $H$ and a regenerated $\acute{H}$ from $\hat{L}$. The difference between regenerated and original images is calculated using the $L1$ function in the following way:
\begin{align}
    L_{C}(\acute{L}, L) &= E_{L}[||\acute{L} - L||_1] \nonumber \\
    L_{C}(\acute{H}, H) &= E_{H}[||\acute{H} - H||_1]
\end{align}

\subsubsection{Identity loss}
To calculate identity loss, we use generators and images from the same domains. More specifically, $H$ should not be modified when it is sent to $G_H$. The same things apply to $L$ and $G_L$. To calculate this loss function, we also use the $L1$ function in the following way:
\begin{align}
    L_{I}(L, G_L(L)) &= E_{L}[||L - G_L(L)||_1] \nonumber \\
    L_{I}(H, G_H(H)) &= E_{H}[||H - G_H(H)||_1]
\end{align}

\subsubsection{Total loss}
The total generator and discriminator loss can be formulated as:
\begin{align}
    L(G_L, G_H, D_L, D_H) &= L_{A}(D_L, L, \hat{L}) + L_{A}(D_H, H, \hat{H}) + \nonumber \\ 
                          &+ \lambda_1 L_{C}(\acute{L}, \hat{L}) + \lambda_1 L_{C}(\acute{H}, \hat{H}) + \nonumber \\
                          &+ \lambda_2 L_{I}(L, G_L(L)) + \lambda_2 L_{I}(H, G_H(H))
\end{align}
\noindent where $\lambda_1$ and $\lambda_2$ are weights of cycle consistency and identity losses. The main problem that discriminators and generators are trying to solve can be mathematically expressed in the following equation:

\begin{equation}
    G_L^{*}, G_H^{*} = arg\min_{G_L, G_H}\max_{D_L, D_H}L(G_L, G_H, D_L, D_H)
\end{equation}

\section{Results and Discussion}

\subsection{Datasets}
In our work, we used two datasets, such as EyeQ \cite{fu2019evaluation} and Mendeley \cite{yoo2020cyclegan}. EyeQ dataset has 16818 "Good", 6434 "Usable", and 5,538 "Bad" images of size 800$\times$800. The Mendeley dataset comprises 1146 "Artifacts" and 1060 "No artifacts" images of size 350$\times$346. For training, we used all "Good," "No artifacts," 5455 "Usable," and 986 "Artifacts" images. The testing set includes 979 "Usable" and 160 "Artifacts" images. To train vessel segmentation model, we used DRIVE \cite{staal2004ridge} dataset that includes 40 images of size 565$\times$584 with corresponding manually annotated segmentation's.  

\subsection{Experimental Setup}
We implement the proposed model in the PyTorch framework. The hardware configurations are Intel Core i7-10700f CPU, 32 GB of RAM, and an NVIDIA GeForce RTX 2080 SUPER 8 GB. The model was set to train for 100 epochs with early stopping as soon as average generator loss stops decreasing. The generator loss has decreased, while discriminator loss has increased for 23 epochs. The generator model improved faster than the discriminator until both converged. 
The total training time is 25 hours for our experiment.


\begin{figure}[!ht]
    \centering
    \includegraphics[width=0.45\textwidth]{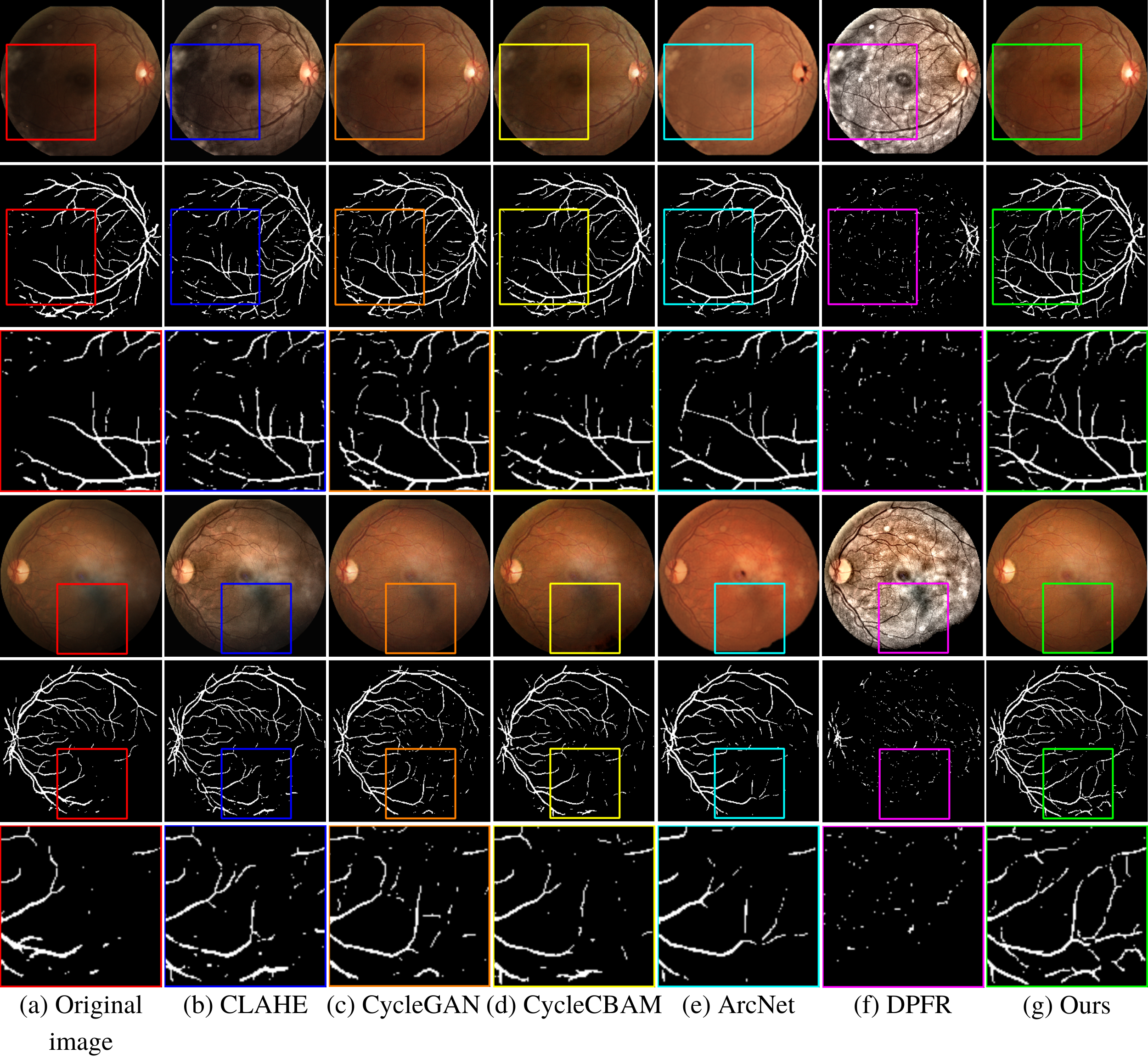}
    \caption{Comparison with state-of-the-art methods. From top to bottom are fundus image, corresponding segmentation and zoomed segmentation. (a) low-quality image, (b) CLAHE \cite{shamsudeen2016enhancement}, (c) CycleGAN \cite{you2019fundus}, (d) CycleCBAM \cite{wan2021retinal}, (e) ArcNet \cite{li2021restoration}, (f) DPFR \cite{zhang2022double}, (g) Ours.}
    \label{fig:qualitative}
\end{figure}

\subsection{Qualitative and Quantitative Performance}
To validate the efficiency of our method, we performed qualitative analysis for restored images using vessel segmentation with 5 state-of-the-art methods: CLAHE \cite{shamsudeen2016enhancement}, CycleGAN \cite{you2019fundus}, Cycle-CBAM \cite{wan2021retinal}, ArcNet \cite{li2021restoration}, DPFR \cite{zhang2022double}. UNet \cite{ronneberger2015u} was trained using pairs of fundus images with corresponding manually segmented images from DRIVE dataset \cite{staal2004ridge}. Fig. \ref{fig:qualitative} shows the restoration and vessel segmentation results. Our method enhanced the retinal image better compared to the state-of-the-art. In addition, our result has less noise compared to CycleGAN \cite{you2019fundus} and CycleCBAM \cite{wan2021retinal}. Algorithm-based methods \cite{shamsudeen2016enhancement, zhang2022double} could improve the contrast making vessels visible, but produced images have artifacts. ArcNet \cite{li2021restoration} failed to restore tiny blood vessels in the dark region.

To further evaluate our results, we conducted the quantitative analysis with the same methods. The evaluation metrics include PSNR, SSIM, and single image test time (SITT), as shown in TABLE \ref{table:quantitative}. As we can see, our method significantly outperforms all other methods in terms of PSNR and SSIM with the EyeQ dataset. Our approach also shows the best results compared to the state-of-the-art for the Mendeley dataset. In terms of SITT, it shows competitive outcomes taking the third position after CLAHE \cite{shamsudeen2016enhancement} and original CycleGAN \cite{you2019fundus}.

\setlength{\tabcolsep}{2pt}
\setlength{\extrarowheight}{1pt}
\begin{table}[!ht]
\caption{Comparison with state-of-the-art methods.}
\label{table:quantitative}
\centering
\begin{tabular}{|c|c|c|c|c|} 
 \hline
 Dataset & Method & PSNR & SSIM & SITT \\ [1ex]
 \hline
 \hline
 \multirow{8}{*}{\specialcell{EyeQ \cite{fu2019evaluation}}} & CLAHE \cite{shamsudeen2016enhancement} & 25.152 dB & 0.716 & \textbf{17ms} \\ [1ex]
 \cline{2-5}
  & CycleGAN \cite{you2019fundus} & 25.577 dB & 0.882 & 90ms \\ [1ex]
  \cline{2-5}
  & Cycle-CBAM \cite{wan2021retinal} & 26.263 dB & 0.901 & 100ms \\ [1ex]
  \cline{2-5}
  & ArcNet \cite{li2021restoration} & 21.073 dB & 0.87 & 183ms \\ [1ex]
  \cline{2-5}
  & DPFR \cite{zhang2022double} & 12.423 dB & 0.34 & 449ms \\ [1ex]
  \cline{2-5}
  & \textbf{Ours} & \textbf{31.138 dB} & \textbf{0.919} & 97ms \\ [1ex]
 \hline
 \hline
 \multirow{8}{*}{\specialcell{Mendeley \cite{yoo2020cyclegan}}} & CLAHE \cite{shamsudeen2016enhancement} & 26.07 dB & 0.77 & \textbf{17ms} \\ [1ex]
 \cline{2-5}
  & CycleGAN \cite{you2019fundus} & 26.08 dB & 0.9 & 90ms \\ [1ex]
  \cline{2-5}
  & Cycle-CBAM \cite{wan2021retinal} & 25.808 dB & 0.901 & 100ms \\ [1ex]
  \cline{2-5}
  & ArcNet \cite{li2021restoration} & 22.17 dB & 0.891 & 183ms \\ [1ex]
  \cline{2-5}
  & DPFR \cite{zhang2022double} & 12.797 dB & 0.304 & 449ms \\ [1ex]
  \cline{2-5}
  & \textbf{Ours} & \textbf{27.798 dB} & \textbf{0.904} & 97ms \\ [1ex]
 \hline
\end{tabular}\vspace{0.1cm}
\end{table} 

\subsection{Ablation study}
\begin{figure}[!ht]
    \centering
    \includegraphics[width=0.49\textwidth]{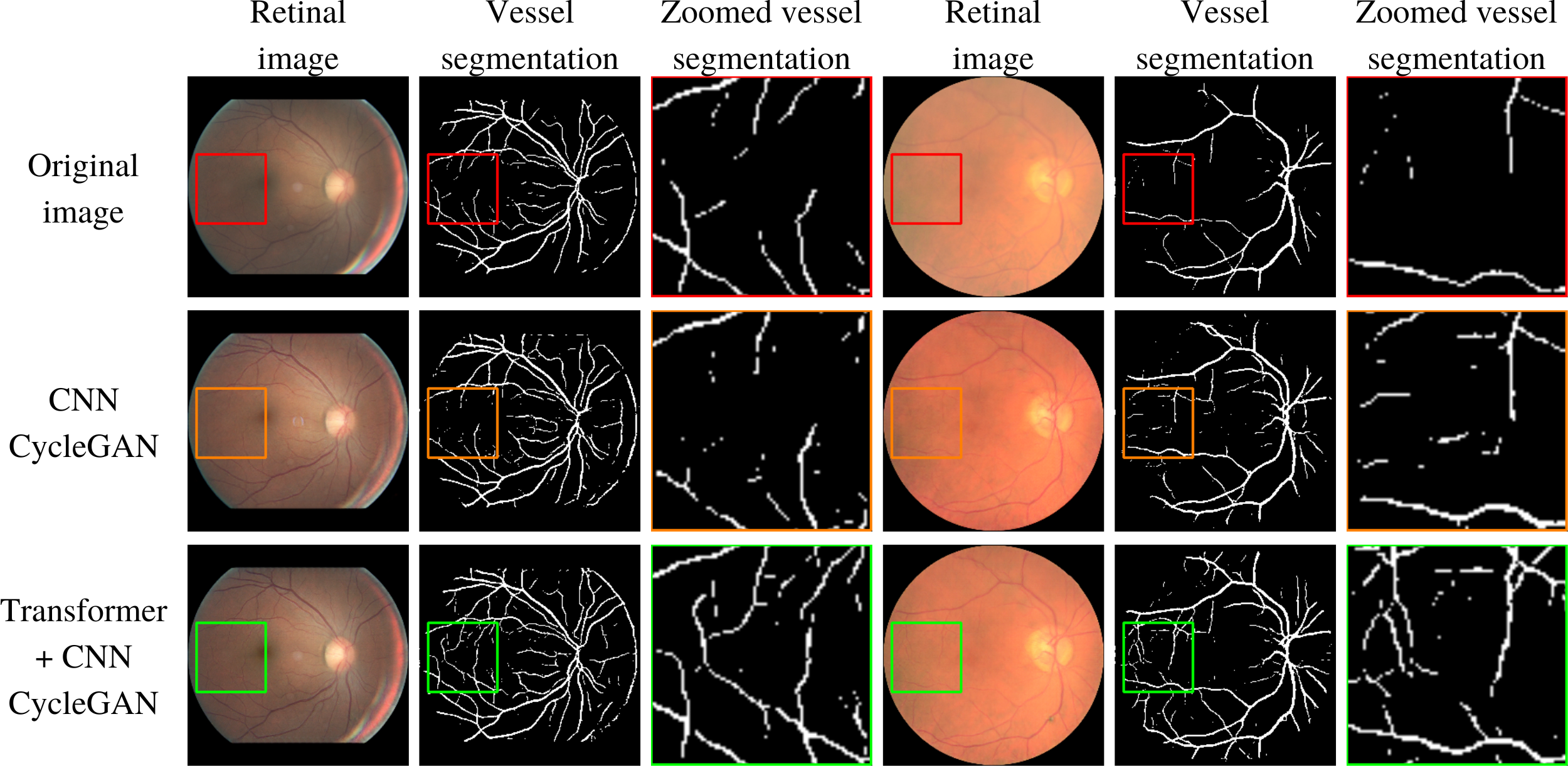}
    \caption{Qualitative comparison of tiny blood vessels restoration.}
    \label{fig:ablation}
\end{figure}

To conduct an ablation study of our work, we trained and tested the original fully convolutional CycleGAN model with the same datasets. Total training time for the traditional CycleGAN required 40 hours to converge, 15 hours more than ours. Additionally, testing PSNR values of our and traditional methods are 31.138 dB and 25.577 dB for the EyeQ dataset, and for the Mendeley dataset, the values are 27.798 dB and 26.08 dB, respectively. The SSIM values are 0.919 and 0.882 for the EyeQ and 0.904 and 0.9 for the Mendeley datasets, respectively. On average, testing one image takes 90ms for CNN CycleGAN and 97ms for our method, meaning that our modification did not add much computational expense. Fig. \ref{fig:ablation} demonstrates the qualitative comparison of these two methods. Our method is better at restoring tiny, barely visible blood vessels and preserving structural and color information of original images.

\section{Conclusion}
In this paper, we combined a vision transformer and convolutional neural network, resulting in better retinal image enhancement than fully convolutional CycleGAN and other state-of-the-art techniques. In addition, our goal was to reduce the computational expense of transformer CycleGAN architecture without compromising the performance. This simple yet effective combination has led to comparatively better quantitatively, qualitatively, and computational performance. The model was tested using two datasets, such as EyeQ \cite{fu2019evaluation} and Mendeley dataset \cite{yoo2020cyclegan}. Testing PSNR is 31.138 dB for the first and 27.798 dB for the second dataset. Testing SSIM is 0.919 and 0.904, respectively. The ablation study demonstrates a significant advantage of our method compared to the original CycleGAN architecture. 




\bibliographystyle{IEEEtran}

\bibliography{strings,refs}
\end{document}